\documentstyle[a4,12pt,epsfig]{article}
\begin{document}
\parskip=5pt plus 1pt minus 1pt

\begin{flushright}
\framebox{\bf hep-ph/0304234}
\end{flushright}

\vspace{0.2cm}

\begin{center}
{\Large\bf Lepton Mass Matrices with Four Texture Zeros}
\end{center}

\vspace{0.3cm}

\begin{center}
{\bf Zhi-zhong Xing}
\footnote{Electronic address: xingzz@mail.ihep.ac.cn} \\
{\it CCAST (World Laboratory), P.O. Box 8730, Beijing 100080, China \\
and Institute of High Energy Physics, Chinese Academy of Sciences, \\
P.O. Box 918 (4), Beijing 100039, China}
\footnote{Mailing address}
\end{center}
\begin{center}
{\bf He Zhang} \\
{\it Physics Department, Jilin University, Changchun 130023, China}
\end{center}

\vspace{1.5cm}

\begin{abstract}
We propose two ans$\rm\ddot{a}$tze of lepton mass matrices with four 
texture zeros, and confront them with current experimental data on
neutrino oscillations. The parameter space of each ansatz is
carefully explored. We find that both ans$\rm\ddot{a}$tze can 
accommodate the normal hierarchy of neutrino masses and the bi-large
pattern of lepton flavor mixing. Their predictions for the effective
mass of the tritium beta decay and that of the neutrinoless double
beta decay are too small to be detectable, but leptonic CP violation 
at the percent level is allowed. Some discussions are also given 
about the seesaw invariance of the four-zero texture of Dirac and
Majorana neutrino mass matrices. 
\end{abstract}

\newpage

\section{Introduction}

The recent KamLAND \cite{KM} and SNO \cite{SNO} experiments have 
provided us with very compelling evidence that the solar neutrino 
deficit is due to the matter-enhanced $\nu_e \rightarrow \nu_\mu$ 
oscillation with a large mixing angle $\theta_{\rm sun} \sim 32^\circ$.
Meanwhile, the K2K \cite{K2K} and Super-Kamiokande \cite{SK} 
experiments have convinced us that the atmospheric neutrino anomaly 
is attributed to the $\nu_\mu \rightarrow \nu_\tau$ oscillation with 
another large mixing angle $\theta_{\rm atm} \sim 45^\circ$. 
In contrast, the non-observation of the
$\overline{\nu}_e \rightarrow \overline{\nu}_e$ oscillation 
in the CHOOZ experiment \cite{CHOOZ} indicates a rather small
(even vanishing) mixing angle $\theta_{\rm chz} < 13^\circ$.
These three mixing angles may straightforwardly be related to
the elements of the $3\times 3$ lepton flavor mixing matrix $V$,
which links the neutrino mass eigenstates $(\nu_1, \nu_2, \nu_3)$ 
to the neutrino flavor eigenstates $(\nu_e, \nu_\mu, \nu_\tau)$. 
To a good degree of accuracy, 
$\theta_{\rm sun}$ and $\theta_{\rm atm}$ describe the flavor 
mixing effect between the 1st and 2nd lepton families and that
between the 2nd and 3rd lepton families, respectively; while the
small mixing angle $\theta_{\rm chz}$ is responsible for the
flavor mixing effect between the 1st and 3rd lepton families.
Thus the lepton flavor mixing matrix $V$ performs a {\it bi-large} 
mixing pattern, which is quite different from the {\it tri-small} 
mixing pattern of the quark flavor mixing matrix \cite{CKM}.

To interpret the bi-large lepton flavor mixing pattern, many 
phenomenological ans$\rm\ddot{a}$tze of lepton mass matrices have 
recently been proposed \cite{Review}. A particularly interesting 
category of the ans$\rm\ddot{a}$tze focus on {\it texture zeros} 
of charged lepton and neutrino mass matrices in a given flavor 
basis
\footnote{For instance, a lot of interest has recently been 
paid to possible two-zero textures of the neutrino mass
matrix in the flavor basis where the charged lepton mass matrix is
diagonal \cite{20}.},
from which some nontrivial relations between flavor mixing angles 
and lepton mass ratios can be derived. The typical example is the
Fritzsch ansatz \cite{Fritzsch} for symmetric lepton mass matrices,
\begin{equation}
M_{l,\nu} \; = \; \left ( \matrix{
{\bf 0}	& ~ C_{l,\nu}	& ~ {\bf 0} \cr
C_{l,\nu}	& ~ {\bf 0}	& ~ B_{l,\nu} \cr
{\bf 0}	& ~ B_{l,\nu}	& ~ A_{l,\nu} \cr} \right ) \; ,
\end{equation}
in which six texture zeros are included 
\footnote{A pair of off-diagonal texture zeros of the charged 
lepton ($M_l$) or neutrino ($M_\nu$) mass matrix have been 
counted, due to symmetry, as one zero \cite{FX03}.}.
It has been shown by one of the authors \cite{X} that this type of 
lepton mass matrices can naturally predict a normal hierarchy of 
neutrino masses and a bi-large pattern of lepton flavor mixing 
angles. Furthermore, Fukugita, Tanimoto and Yanagida \cite{FTY}
have demonstrated that very similar phenomenological predictions 
can also be achieved from a simple but interesting ansatz of 
lepton mass matrices based on both the Fritzsch texture  
and the seesaw mechanism \cite{SS}. 

The present paper aims to analyze the generalized Fritzsch ansatz 
of lepton mass matrices with four texture zeros,
\begin{equation}
M_{l,\nu} \; = \; \left ( \matrix{
{\bf 0}	& ~ C_{l,\nu}	& ~ {\bf 0} \cr
C_{l,\nu}	& ~ \tilde{B}_{l,\nu}	& ~ B_{l,\nu} \cr
{\bf 0}	& ~ B_{l,\nu}	& ~ A_{l,\nu} \cr} \right ) \; ,
\end{equation} 
and its consequences on the neutrino mass spectrum, flavor mixing
and CP violation. It is well known that the four-zero texture of
quark mass matrices is more successful than the six-zero texture of
quark mass matrices to interpret the strong hierarchy of quark 
masses and the smallness of flavor mixing angles. 
The spirit of lepton-quark similarity motivates us to conjecture
that the lepton mass matrices might have the same texture zeros
as the quark mass matrices. Such a conjecture is indeed reasonable
in some specific models of grand unified theories \cite{Review}, 
in which the
mass matrices of leptons and quarks are related to each other by
a new kind of flavor symmetry. That is why the four-zero texture
of lepton mass matrices has been considered as a typical
example in some model-building works \cite{40}. However, a careful 
and complete analysis of its phenomenological implications has not 
been done in the literature.

Naively, there is no doubt that the four-zero texture of $M_{l,\nu}$ 
in Eq. (2), which has more free parameters than the Fritzsch 
texture of $M_{l,\nu}$ in Eq. (1), must be able to interpret 
the observed bi-large pattern of lepton flavor mixing. To improve
the analytical calculability and numerical
predictability, one may follow a realistic strategy to concentrate 
on part of the whole parameter space of the four-zero texture of
lepton mass matrices. The same strategy has actually been adopted 
in the study of the four-zero texture of quark mass 
matrices \cite{FX99}. In this paper, we shall consider two 
simplified versions of $M_l$ and $M_\nu$ given in Eq. (2):
\begin{enumerate}
\item	$|\tilde{B}_l| = m_\mu$ and $|\tilde{B}_\nu| = m_2$, where
$m_\mu$ and $m_2$ stand respectively for the physical masses of 
$\mu$ and $\nu_2$. This interesting case, to be referred to as
Ansatz (A), has been briefly discussed in Ref. \cite{40}. 
\item	$|\tilde{B}_l| = |B_l|$ and $|\tilde{B}_\nu| = |A_\nu|$. 
This specific case, to be referred to as Ansatz (B), 
has not been discussed in the literature.
\end{enumerate}
Our main purpose is to explore the allowed parameter space of each
ansatz and its implications on the neutrino mass spectrum and lepton
flavor mixing measurables. We find that both ans$\rm\ddot{a}$tze
are favored by current experimental data on neutrino oscillations.
Their predictions for the effective mass of the tritium beta decay 
and that of the neutrinoless double beta decay are too small to be 
detectable, but leptonic CP violation at the percent level is 
definitely allowed. Finally, we present some brief discussions about 
the seesaw invariance of the four-zero texture of Dirac and
Majorana neutrino mass matrices. 
  
\section{Framework}

Note that all non-zero elements of $M_l$ and $M_\nu$ in Eq. (2)
are in general complex. If the condition
\begin{equation}
\arg (A_{l,\nu}) + \arg (\tilde{B}_{l,\nu})  = 
2 \arg (B_{l,\nu}) \;
\end{equation}
is satisfied, then $M_{l,\nu}$ can be decomposed as
\begin{equation}
M_l  = P^T_l \overline{M}_l P_l \; , ~~
M_\nu = P^T_\nu \overline{M}_\nu P_\nu \; ,
\end{equation}
where
\begin{eqnarray}
\overline{M}_{l,\nu} & = & \left ( \matrix{
{\bf 0}  	& |C_{l,\nu}|   & {\bf 0} \cr
|C_{l,\nu}|  	& |\tilde{B}_{l,\nu}|   	& |B_{l,\nu}| \cr
{\bf 0}  	& |B_{l,\nu}|  	& |A_{l,\nu}| \cr} \right ) \; ,
\nonumber \\
P_{l,\nu} & = & \left ( \matrix{
e^{i\alpha^{~}_{l,\nu}} & 0 & 0 \cr
0 & ~ e^{i\beta^{~}_{l,\nu}} ~ & 0 \cr
0 & 0 & e^{i\gamma^{~}_{l,\nu}} \cr} \right ) \; ,
\end{eqnarray}
with $\arg (A_{l,\nu}) = 2\gamma^{~}_{l,\nu}$,
$\arg (B_{l,\nu}) = \beta^{~}_{l,\nu} + \gamma^{~}_{l,\nu}$,
$\arg(\tilde{B}_{l,\nu}) = 2\beta^{~}_{l,\nu}$ and
$\arg (C_{l,\nu}) = \alpha^{~}_{l,\nu} + \beta^{~}_{l,\nu}$. 
The real symmetric matrices
$\overline{M}_l$ and $\overline{M}_\nu$ 
can be diagonalized by use of the following unitary transformations:
\begin{eqnarray} 
U^T_l \overline{M}_l U_l & = & \left ( \matrix{
m_e & 0 & 0 \cr
0 & ~ m_\mu ~ & 0 \cr
0 & 0 & m_\tau \cr} \right ) \; , 
\nonumber \\
U^T_\nu \overline{M}_\nu U_\nu & = & \left ( \matrix{
m_1 & 0 & 0 \cr
0 & ~ m_2 ~ & 0 \cr
0 & 0 & m_3 \cr} \right ) \; , 
\end{eqnarray}
in which $(m_e, m_\mu, m_\tau)$ and $(m_1, m_2, m_3)$ denote the 
physical masses of charged leptons and neutrinos, respectively.   
The lepton flavor mixing matrix $V$ arises from the mismatch between 
the diagonalization of the charged lepton mass matrix $M_l$ and that 
of the neutrino mass matrix $M_\nu$. Therefore, we obtain 
$V = U^T_l \left (P^*_l P_\nu \right ) U^*_\nu$, whose nine matrix
elements read explicitly as
\begin{equation}
V_{pq} \; = \; U^l_{1 p} U^{\nu *}_{1 q} e^{i\alpha} ~ + ~ 
U^l_{2 p} U^{\nu *}_{2 q} e^{i \beta} ~ + ~ 
U^l_{3 p} U^{\nu *}_{3 q} e^{i \gamma} \; ,
\end{equation}
where the subscripts $p$ and $q$ run respectively over $(e, \mu, \tau)$
and $(1,2,3)$, and the parameters $\alpha$, $\beta$ and $\gamma$ are 
defined as
\begin{equation}
\alpha \equiv \alpha^{~}_\nu - \alpha^{~}_l \; , ~~~
\beta \equiv \beta^{~}_\nu - \beta^{~}_l \; , ~~~
\gamma \equiv \gamma^{~}_\nu - \gamma^{~}_l \; .
\end{equation}
Note that the overall phase of $V$ has nothing to do with the 
experimental observables. Hence only two combinations of three
phases $(\alpha, \beta, \gamma)$ are physically relevant. For
simplicity, we take $\gamma =0$ in the following.

The matrix elements of $V$ depend both on the ratios of lepton
masses, 
\begin{equation}
x^{~}_l \equiv \frac{m_e}{m_\mu} \; , ~~
y^{~}_l \equiv \frac{m_\mu}{m_\tau} \; , ~~
x_\nu \equiv \frac{m_1}{m_2} \; , ~~
y_\nu \equiv \frac{m_2}{m_3} \; ,
\end{equation}
and on the phase parameters $\alpha$ and $\beta$. As the values of 
$m_e$, $m_\mu$ and $m_\tau$ have precisely been measured \cite{PDG},
we have $x^{~}_l \approx 0.00484$ and $y^{~}_l \approx 0.0594$ to
a good degree of accuracy. The other four free parameters can be
determined or constrained from the present experimental data on 
neutrino oscillations. As the neutrino mass-squared differences of 
solar and atmospheric neutrino oscillations are given by
\begin{eqnarray}
\Delta m^2_{\rm sun} & \equiv & \left | m^2_2 - m^2_1 \right | 
= m^2_2 \left | 1 - x^2_\nu \right | \; ,
\nonumber \\
\Delta m^2_{\rm atm} & \equiv & \left | m^2_3 - m^2_2 \right |
= m^2_3 \left | 1 - y^2_\nu \right | \; ,
\end{eqnarray}
the observed hierarchy $\Delta m^2_{\rm sun} \ll \Delta m^2_{\rm atm}$
may impose a very strong constraint on the values of $(x_\nu, y_\nu)$:
\begin{equation}
R_\nu \; \equiv \; \frac{\Delta m^2_{\rm sun}}{\Delta m^2_{\rm atm}}
\; =\; y^2_\nu \left | \frac{1 - x^2_\nu}{1 - y^2_\nu} 
\right | \ll 1 \; .
\end{equation}
On the other hand, the mixing factors of solar, atmospheric and
CHOOZ reactor neutrino oscillations are related to the matrix elements 
of $V$ in the following way:
\begin{eqnarray}
\sin^2 2 \theta_{\rm sun} & = & 4 |V_{e1}|^2 |V_{e2}|^2 \; ,
\nonumber \\
\sin^2 2 \theta_{\rm atm} & = & 4 |V_{\mu 3}|^2 
\left (1 - |V_{\mu 3}|^2 \right ) \; ,
\nonumber \\
\sin^2 2\theta_{\rm chz} & = & 4 |V_{e3}|^2
\left ( 1 - |V_{e3}|^2 \right ) \; .
\end{eqnarray}
In view of the KamLAND \cite{KM}, SNO \cite{SNO}, 
K2K \cite{K2K}, Super-Kamiokande \cite{SK} and CHOOZ \cite{CHOOZ}
data on neutrino oscillations, we have 
$\Delta m^2_{\rm sun} \in [5.9, ~ 8.8] \times 10^{-5} ~ {\rm eV}^2$,
$\sin^2 \theta_{\rm sun} \in [0.25, ~ 0.40]$ \cite{Fit};
$\Delta m^2_{\rm atm} \in [1.65, ~ 3.25] \times 10^{-3} ~ {\rm eV}^2$,
$\sin^2 2\theta_{\rm atm} \in [0.88, ~ 1.00]$ \cite{Fogli}; and
$\sin^2 2\theta_{\rm chz} < 0.2$ at the $90\%$ confidence level.
With the help of these experimental results, the allowed ranges 
of $x_\nu$, $y_\nu$, $\alpha$ and $\beta$ can be obtained from 
Eqs. (11) and (12).

Once the values of $x_\nu$ and $y_\nu$ are determined or 
constrained from current experimental data, we are able to calculate 
the absolute values of three neutrino masses by use of Eq. (10):
\begin{eqnarray}
m_3 & = & \frac{1}{\sqrt{|1 - y^2_\nu|}} 
\sqrt{\Delta m^2_{\rm atm}} \;\; ,
\nonumber \\
m_2 & = & \frac{y_\nu}{\sqrt{|1 - y^2_\nu|}} 
\sqrt{\Delta m^2_{\rm atm}} 
\nonumber \\
& = & \frac{1}{\sqrt{|1 - x^2_\nu|}} 
\sqrt{\Delta m^2_{\rm sun}} \;\; ,
\nonumber \\
m_1 & = & \frac{x_\nu}{\sqrt{|1 - x^2_\nu|}} 
\sqrt{\Delta m^2_{\rm sun}} \;\; .
\end{eqnarray}
In addition, interesting predictions can be achieved for the 
effective mass of the tritium beta decay $\langle m\rangle_e$
and that of the neutrinoless double beta decay $\langle m\rangle_{ee}$:
\begin{eqnarray}
\langle m\rangle^2_e & \equiv & \sum^3_{i=1} 
\left ( m^2_i |V_{ei}|^2 \right )
\nonumber \\
& = & m^2_3 \left (x^2_\nu y^2_\nu |V_{e1}|^2 + y^2_\nu |V_{e2}|^2 + 
|V_{e3}|^2 \right ) \; ,
\nonumber \\
\langle m\rangle_{ee} & \equiv & \left | \sum^3_{i=1} 
\left ( m_i V^2_{ei} \right ) \right |
\nonumber \\
& = & m_3 \left | x_\nu y_\nu V^2_{e1} + y_\nu V^2_{e2} + 
V^2_{e3} \right | \; .
\end{eqnarray}
The present experimental upper bound on $\langle m\rangle_e$ is
$\langle m\rangle_e < 3$ eV \cite{PDG}, while the sensitivity of the 
proposed KATRIN experiment is expected to reach 
$\langle m\rangle_e \sim 0.3$ eV \cite{K}. In comparison, 
the upper limit $\langle m\rangle_{ee} < 0.35$ eV has been set by the 
Heidelberg-Moscow Collaboration \cite{HM} at the $90\%$ confidence level
\footnote{If the reported evidence for the existence of the neutrinoless
double beta decay \cite{KK} is taken into account, one has 
$0.05 ~ {\rm eV} \leq \langle m\rangle_{ee} \leq 0.84 ~ {\rm eV}$ at
the $95\%$ confidence level.}.
The sensitivity of the next-generation experiments
for the neutrinoless double beta decay is possible to reach
$\langle m\rangle_{ee} \sim 10$ meV to 50 meV \cite{B}.

The strength of CP violation in neutrino oscillations, which is 
measured by the Jarlskog invariant $\cal J$ \cite{Jarlskog},
can also be predicted from the four-zero texture of lepton mass 
matrices under consideration. Indeed, $\cal J$ is defined through the 
following equation:
\begin{equation}
{\rm Im} \left ( V_{a i} V_{b j} V^*_{a j} V^*_{b i} \right ) \; =\; 
{\cal J} \sum_{c, k} 
\left ( \epsilon_{a b c} \epsilon_{ijk} \right ) \; ,
\end{equation}
where the subscripts $(a, b, c)$ and $(i, j, k)$ run respectively over 
$(e, \mu, \tau)$ and $(1,2,3)$. The magnitude of $\cal J$ depends 
both on $(x_\nu, y_\nu)$ and on $(\alpha, \beta)$. If 
$|{\cal J}| \sim 1\%$ is achievable, then leptonic CP- and T-violating 
effects could be measured in a variety of long-baseline neutrino 
oscillation experiments \cite{LBL} in the future.

\section{Ansatz (A)}

Now let us consider Ansatz (A), in which the requirements
$|\tilde{B}_l| = m_\mu$ and $|\tilde{B}_\nu| = m_2$ are imposed
on $M_l$ and $M_\nu$ in Eq. (2). Similar conditions  
($|\tilde{B}_{\rm u}| = m_c$ and $|\tilde{B}_{\rm d}| = m_s$)
have actually been taken in some literature \cite{40} for the 
four-zero texture of quark mass matrices. Following Eqs. (3) and (4), 
we factor out the complex phases of $M_{l,\nu}$. Then three
elements of the real symmetric mass matrix $\overline{M}_{l,\nu}$ 
can be expressed in terms of its three mass eigenvalues:
\begin{eqnarray}
|A_l| & = & m_\tau - m_e \; , 
\nonumber \\
|B_l| & = & \left [ \frac{m_e m_\tau (m_\tau - m_e - m_\mu)}
{m_\tau - m_e} \right ]^{1/2} \; ,
\nonumber \\
|C_l| & = & \left ( \frac{m_e m_\mu m_\tau}{m_\tau - m_e} 
\right )^{1/2} \; ;
\end{eqnarray}
and
\begin{eqnarray}
|A_\nu| & = & m_3 - m_1 \; , 
\nonumber \\
|B_\nu| & = & \left [ \frac{m_1 m_3 (m_3 - m_1 - m_2)}
{m_3 - m_1} \right ]^{1/2} \; ,
\nonumber \\
|C_\nu| & = & \left ( \frac{m_1 m_2 m_3}{m_3 - m_1} \right )^{1/2} \; .
\end{eqnarray}
The elements of the unitary transformation matrix $U_{l,\nu}$,
which is used to diagonalize $\overline{M}_{l,\nu}$ in Eq. (6), can in 
turn be expressed in terms of the ratios $x^{~}_{l,\nu}$ and
$y^{~}_{l,\nu}$ as follows (the indices ``$l$'' and ``$\nu$'' are 
neglected for simplicity):
\begin{eqnarray}
U_{11} & = & + i \left [ \frac{1}{(1+x)(1-x^2y^2)} \right ]^{1/2} \; , 
\nonumber \\
U_{12} & = & + \left [ \frac{x(1-y-xy)}{(1+x)(1-y)(1-xy)} 
\right ]^{1/2} \; ,
\nonumber \\
U_{13} & = & + \left [ \frac{x^2y^3}{(1-y)(1-x^2y^2)} 
\right ]^{1/2} \; , 
\nonumber \\
U_{21} & = & - i \left [ \frac{x}{(1+x)(1+xy)} \right ]^{1/2} \; ,
\nonumber \\
U_{22} & = & + \left [ \frac{1-y-xy}{(1+x)(1-y)} \right ]^{1/2} \; , 
\nonumber \\
U_{23} & = & + \left [ \frac{xy}{(1-y)(1+xy)} \right ]^{1/2} \; ,
\nonumber \\
U_{31} & = & + i \left [ \frac{x^2y(1-y-xy)}{(1+x)(1-x^2y^2)} 
\right ]^{1/2} \; , 
\nonumber \\
U_{32} & = & - \left [ \frac{xy}{(1+x)(1-y)(1-xy)} \right ]^{1/2} \; , 
\nonumber \\
U_{33} & = & + \left [ \frac{1-y-xy}{(1-y)(1-x^2y^2)} \right ]^{1/2} \; .
\end{eqnarray}
Note that $U_{i1}$ (for $i=1,2,3$) are imaginary, and their nontrivial
phases are due to the negative determinant of $\overline{M}_{l,\nu}$.

The four free parameters $x_\nu$, $y_\nu$, $\alpha$ and $\beta$ can
be constrained by use of Eqs. (11) and (12) as well as current data
on neutrino oscillations. Their allowed ranges are shown in Fig. 1.
We see that $x_\nu \sim 0.86$ and $y_\nu \sim 0.35$ typically hold.
Thus the neutrino mass spectrum satisfies $m_1 < m_2 < m_3$. For
$\alpha$ and $\beta$ varying from 0 to $2\pi$, we find that about 
half of the whole $(\alpha, \beta)$ parameter space can be excluded.
Note that the correlation between $\alpha$ and $\beta$ in Ansatz (A) 
is not as strong as the phase correlation in the Fritzsch 
ansatz of lepton mass matrices \cite{X,FTY}. The reason is simply
that the contribution of $M_l$ to $V$ is much smaller in Ansatz (A) 
than in the Fritzsch ansatz. Hence the relative phases between
$M_l$ and $M_\nu$ in the former cannot significantly affect the 
magnitudes of nine matrix elements of $V$.

Fig. 1 also shows the ouputs of $\sin^2 2\theta_{\rm atm}$ versus
$\sin^2\theta_{\rm sun}$ and $\sin^2 2\theta_{\rm chz}$ versus
$R_\nu$ restricted by Ansatz (A). It can be seen that larger
values of $\sin^2\theta_{\rm sun}$ roughly correspond to smaller
values of $\sin^2 2\theta_{\rm atm}$. In addition, the ansatz
predicts $\sin^2 2\theta_{\rm chz} \geq 0.08$, a lower bound
which is easily accessible in the upcoming long-baseline neutrino
oscillation experiments \cite{LBL}. If the upper limit
$\sin^2 2\theta_{\rm chz} < 0.1$ instead of 
$\sin^2 2\theta_{\rm chz} < 0.2$ is input in the numerical 
calculations, one will arrive at 
$\sin^2 2\theta_{\rm atm} \leq 0.91$. Such a low value of
$\sin^2 2\theta_{\rm atm}$ is tolerable, but not favored by
current data. It becomes clear that
the mixing angles $\theta_{\rm sun}$, $\theta_{\rm atm}$ and
$\theta_{\rm chz}$ are strongly correlated with one another in
Ansatz (A). Thus more precise data on three mixing angles may
provide a sensitive test of this phenomenological scenario. 

The result $y^2_\nu \sim 0.1$ implies that 
$m_3 \approx \sqrt{\Delta m^2_{\rm atm}}$ is an acceptable 
approximation. More exactly, we obtain 
$m_3 \approx (4.3 - 6.1) \times 10^{-2}$ eV, 
$m_2 \approx (1.4 - 2.3) \times 10^{-2}$ eV and 
$m_1 \approx (1.1 - 2.1) \times 10^{-2}$ eV from Eq. (13).
In calculating the allowed ranges of $m_1$ and $m_2$, we have
ignored their correlation induced by the model itself. This
generous treatment has no conflict with the plot of 
$(x_\nu, y_\nu)$ in Fig. 1, in which $x_\nu < 1$ results from the 
correlation between $m_1/m_2$ and $m_2/m_3$.
The sum of three neutrino masses is consistent with
$m_1 + m_2 + m_3 < 0.71$ eV, an upper bound set by the recent 
WMAP data \cite{WMAP}. We compute the effective mass of the
tritium beta decay and that of the neutrinoless double beta decay
by use of Eq. (14), and present the numerical results in Fig. 1.
One can see that $\langle m\rangle_e \sim 10^{-2}$ eV and 
$\langle m\rangle_{ee} \sim 10^{-3}$ eV typically hold.
Both quantities are too small to be measured in practice. 
Similarly, there is no hope to detect the effective (kinematic)
masses of muon and tau neutrinos \cite{NO}. The numerical results
for the Jarlskog parameter $\cal J$ and the 
smallest matrix element $|V_{e3}|$ are also shown in 
Fig. 1. We find that the magnitude of ${\cal J}$ may nearly be
$1\%$, if $|V_{e3}|$ is close to its upper bound. It is possible 
to measure leptonic CP violation of this order in the future 
neutrino factories, if the terrestrial matter effects can be 
under control.

Finally we illustrate the typical texture of $\overline{M}_{l,\nu}$ 
by taking $x_\nu \approx 0.86$, $y_\nu \approx 0.35$ and
$m_3 \approx 0.05$ eV. The result is
\begin{eqnarray}
\overline{M}_l & \approx & 1.78 ~ {\rm GeV} \times 
\left ( \matrix{
{\bf 0}  & 0.0041   & {\bf 0} \cr
0.0041 	& 0.059   & 0.016 \cr
{\bf 0}  & 0.016 & {\bf 1} \cr} \right ) \; ,
\nonumber \\
\overline{M}_\nu & \approx & 3.50 \times 10^{-2} ~ {\rm eV} \times 
\left ( \matrix{
{\bf 0}  & 0.56   & {\bf 0} \cr
0.56  	& 0.50   & 0.55 \cr
{\bf 0}  & 0.55 & {\bf 1} \cr} \right ) \; .
\end{eqnarray}
It becomes obvious that lepton flavor mixing is dominated by the 
neutrino sector, as the matrix elements of $\overline{M}_l$ have
a very strong hierarchy.

\section{Ansatz (B)}

We proceed to consider Ansatz (B), in which the requirements 
$|\tilde{B}_l| = |B_l|$ and $|\tilde{B}_\nu| = |A_\nu|$ are imposed 
on $M_l$ and $M_\nu$ in Eq. (2). Note that the condition
$|\tilde{B}_l| = |B_l|$ is similar to 
$|\tilde{B}_{\rm u}| \approx |B_{\rm u}|$ and 
$|\tilde{B}_{\rm d}| \approx |B_{\rm d}|$ for the four-zero texture
of quark mass matrices \cite{FX99}, in view of the fact that
charged leptons have a strong mass hierarchy as quarks. Because
the condition $|\tilde{B}_l| = |B_l|$ leads to $|B_l| \approx m_\mu$ 
in the leading-order approximation, it is essentially equivalent to
the condition $|\tilde{B}_l| = m_\mu$ taken in Ansatz (A). In contrast,
the requirement $|\tilde{B}_\nu| = |A_\nu|$ is motivated by the
experimental fact that the mixing angle of atmospheric neutrino
oscillations is about $45^\circ$ \cite{SK} (namely, the diagonalization
of the (2,3) subsector of $M_\nu$ may give rise to a rotation angle of
$45^\circ$, if the condition $|\tilde{B}_\nu| = |A_\nu|$ is satisfied).
Such a phenomenological 
hypothesis for the texture of $M_\nu$ results in an apparent 
``structural asymmetry'' between $M_l$ and $M_\nu$, but its 
consequences on lepton flavor mixing are simple and interesting.
Following Eqs. (3) and (4),
we can factor out the complex phases of $M_{l,\nu}$. Although 
we are able to exactly express both $(|A_l|, |B_l|, |C_l|)$ and  
$(|A_\nu|, |B_\nu|, |C_\nu|)$ in terms of the corresponding mass
eigenvalues, the formulas for the former are too complicated to be
instructive. It is therefore better to make some analytical
approximations in deriving $|A_l|$, $|B_l|$ and $|C_l|$. In view of
the strong mass hierarchy in the charged lepton sector, we expect
that $|A_l| \gg |B_l| \gg |C_l|$ naturally holds. Then we obtain
\begin{eqnarray}
|A_l| & \approx & m_\tau 
\left (1 - \frac{m^2_\mu}{m^2_\tau} \right ) \; ,
\nonumber \\
|B_l| & \approx & m_\mu 
\left ( 1 + \frac{m_\mu}{m_\tau} \right ) \; ,
\nonumber \\
|C_l| & \approx & \sqrt{m_e m_\mu} 
\left (1 + \frac{m^2_\mu}{2m^2_\tau} \right ) \; ,
\end{eqnarray}
to a good degree of accuracy. In contrast, the expressions of
$|A_\nu|$, $|B_\nu|$ and $|C_\nu|$ are exact:
\begin{eqnarray}
|A_\nu| & = & \frac{m_3+m_2-m_1}{2} \; ,
\nonumber \\
|B_\nu| & = & \frac{1}{2} \left[ \frac{\left (m_3+m_2+m_1 \right ) 
\left (m_3-m_2-m_1 \right ) \left (m_3-m_2+m_1 \right )}
{m_3+m_2-m_1}\right]^{1/2} \; ,
\nonumber \\
|C_\nu| & = & \left (\frac{2 m_1 m_2 m_3}{m_3+m_2-m_1} \right)^{1/2} \; .
\end{eqnarray}
The unitary transformation matrix $U_l$, which has been defined to 
diagonalize $\overline{M}_l$ in Eq. (6), can approximately be given as
\begin{equation}
U_l \; \approx \; \left ( \matrix{ 
\displaystyle i \left (1 - \frac{x^{~}_l}{2} \right ) & 
\displaystyle \sqrt{x^{~}_l} &
\displaystyle y^{~}_l \sqrt{x^{~}_l y^{~}_l} \cr
\displaystyle -i \sqrt{x^{~}_l} & 
\displaystyle ~~ 1 - \frac{x^{~}_l}{2} - \frac{y^2_l}{2} ~~ & 
\displaystyle y^{~}_l \cr
\displaystyle i y^{~}_l \sqrt{x^{~}_l} & 
\displaystyle - y^{~}_l & 
\displaystyle 1 - \frac{y^2_l}{2} \cr} \right ) \; ,
\end{equation}
where the fact of $x^{~}_l \sim y^2_l$ has been taken into account.
In addition, the matrix elements of $U_\nu$ can be expressed in terms 
of $x_\nu$ and $y_\nu$ as follows:
\begin{eqnarray}
U_{11}^\nu & = & +i \left [\frac{(1+y_{\nu}+x_{\nu}y_{\nu})}
{(1+x_{\nu})(1+x_{\nu}y_{\nu})(1+y_{\nu}-x_{\nu}y_{\nu})} \right ]^{1/2} \; ,
\nonumber \\
U_{12}^\nu & = & +\left [\frac{x_{\nu}(1-y_{\nu}-x_{\nu}y_{\nu})}{(1+x_{\nu})
(1-y_{\nu})(1+y_{\nu}-x_{\nu}y_{\nu})}\right]^{1/2} \; ,
\nonumber \\
U_{13}^\nu & = & +\left [\frac{x_{\nu}y_{\nu}^2(1-y_{\nu}+x_{\nu}y_{\nu})}
{(1-y_{\nu})(1+x_{\nu}y_{\nu})(1+y_{\nu}-x_{\nu}y_{\nu})} \right ]^{1/2} \; ,
\nonumber \\
U_{21}^\nu & = & -i\left[\frac{x_{\nu}(1+y_{\nu}+x_{\nu}y_{\nu})}
{2(1+x_{\nu})(1+x_{\nu}y_{\nu})} \right ]^{1/2} \; ,
\nonumber \\
U_{22}^\nu & = & +\left [\frac{(1-y_{\nu}-x_{\nu}y_{\nu})}
{2(1+x_{\nu})(1-y_{\nu})} \right ]^{1/2} \; ,
\nonumber \\
U_{23}^\nu & = & +\left [\frac{(1-y_{\nu}+x_{\nu}y_{\nu})}
{2(1-y_{\nu})(1+x_{\nu}y_{\nu})} \right ]^{1/2} \; ,
\nonumber \\
U_{31}^\nu & = & +i\left [\frac{x_{\nu}(1-y_{\nu}-x_{\nu}y_{\nu})
(1-y_{\nu}+x_{\nu}y_{\nu})}{2(1+x_{\nu})(1+x_{\nu}y_{\nu})
(1+y_{\nu}-x_{\nu}y_{\nu})} \right ]^{1/2} \; ,
\nonumber \\
U_{32}^\nu & = & -\left [\frac{(1+y_{\nu}+x_{\nu}y_{\nu})
(1-y_{\nu}+x_{\nu}y_{\nu})}{2(1+x_{\nu})(1-y_{\nu})
(1+y_{\nu}-x_{\nu}y_{\nu})} \right ]^{1/2} \; ,
\nonumber \\
U_{33}^\nu & = & +\left [\frac{(1+y_{\nu}+x_{\nu}y_{\nu})
(1-y_{\nu}-x_{\nu}y_{\nu})}{2(1-y_{\nu})(1+x_{\nu}y_{\nu})
(1+y_{\nu}-x_{\nu}y_{\nu})} \right ]^{1/2} \; .
\end{eqnarray}
Note again that $U_{i1}^l$ and $U_{i1}^{\nu}$ (for $i=1,2,3$) are 
imaginary, and their nontrivial phases are due to 
${\rm Det}(\overline{M}_l) <0$ and ${\rm Det}(\overline{M}_\nu) <0$.

With the help of Eqs. (11) and (12), one may compute the ranges of 
$x_{\nu}$, $y_{\nu}$, $\alpha$ and $\beta$ allowed by current 
data on neutrino oscillations. The numerical results are shown in
Fig. 2. We see that the allowed range of $x_\nu$ in Ansatz (B) is much 
larger than that in Ansatz (A). As both $x_\nu <1$ and $y_\nu <1$ hold,
the neutrino mass spectrum satisfies $m_1 < m_2 < m_3$.
Note that there is little restriction on the phase parameters $\alpha$ 
and $\beta$. This feature can easily be understood: the contribution of 
$M_\nu$ to $V$ is dominant over that of $M_l$ to $V$, thus the 
magnitudes of nine matrix elements of $V$ are essentially insensitive 
to the relative phases between $M_l$ and $M_{\nu}$.

The outputs of $\sin^2 2\theta_{\rm atm}$ versus 
$\sin^2 \theta_{\rm sun}$ and $\sin^2 2\theta_{\rm chz}$ versus $R_\nu$ 
are also shown in Fig. 2. We see that 
$\sin^2 2\theta_{\rm atm} \geq 0.94$ holds
and $\sin^2 2\theta_{\rm atm} \approx 1$ is particularly favored.
The latter is a natural consequence of the specific texture of 
$M_\nu$, whose (2,3) subsector can be diagonalized by a rotation of 
$45^\circ$. Thus $|V_{\mu 3}| \approx 1/\sqrt{2}$ leads to
$\sin^2 2\theta_{\rm atm} \approx 1$ in Ansatz (B). Another feature
of Ansatz (B) is that changing the upper bound of 
$\sin^2 2\theta_{\rm chz}$ from 0.2 to 0.1 does not significantly 
affect the allowed range of $\sin^2 2\theta_{\rm atm}$.
However, the correlation between $\sin^2 2\theta_{\rm chz}$ and
$R_\nu$ is stronger in Ansatz (B) than in Ansatz (A).

The approximation $m_3 \approx \sqrt{\Delta m^2_{\rm atm}}$ is 
reasonably good in Ansatz (B). Numerically, we obtain 
$m_3 \approx (4.2 - 5.8) \times 10^{-2} ~ {\rm eV}$,
$m_2 \approx (0.8 - 1.4) \times 10^{-2} ~ {\rm eV}$ and
$m_1 \approx (0.3 - 1.1) \times 10^{-2} ~ {\rm eV}$ from Eq. (13).
This neutrino mass spectrum is quite similar to that in Ansatz (A).
The results for the effective mass of the tritium beta 
decay and that of the neutrinoless double beta decay are also
illustrated in Fig. 2. One can see that 
$\langle m \rangle_e \sim 10^{-2} ~ {\rm eV}$ and 
$\langle m \rangle_{ee} \sim 10^{-3} ~ {\rm eV}$ hold in Ansatz (B).
Both of them are too small to be detected in reality. In addition,
Fig. 2 shows that the magnitude of the Jarlskog parameter
$\cal J$ may nearly be $1.5\%$, if $|V_{e3}|$ is larger than 0.15. 
This result implies that it is possible to observe leptonic CP 
violation in the future long-baseline neutrino oscillation experiments.

To illustrate the texture of $\overline{M}_{l,\nu}$, we typically take
$x_{\nu} \approx 0.6$, $y_{\nu} \approx 0.21$ and 
$m_3 \approx 0.05 ~{\rm eV}$. The numerical result is 
\begin{eqnarray}
\overline{M}_l & \approx & 1.77 ~ {\rm GeV} \times 
\left ( \matrix{
{\bf 0}  & 0.0042   & {\bf 0} \cr
0.0042 	& 0.064   & 0.064 \cr
{\bf 0}  & 0.064 & {\bf 1} \cr} \right ) \; ,
\nonumber \\
\overline{M}_\nu & \approx & 2.71 \times 10^{-2} ~ {\rm eV} \times 
\left ( \matrix{
{\bf 0}  & 0.41   & {\bf 0} \cr
0.41  	& {\bf 1}   & 0.80 \cr
{\bf 0}  & 0.80 & {\bf 1} \cr} \right ) \; .
\end{eqnarray}
The similarity and difference between two ans$\rm\ddot{a}$tze are
therefore obvious.

\section{Seesaw}

Two ans$\rm\ddot{a}$tze of lepton mass matrices discussed above
can self-consistently describe the observed features of lepton flavor 
mixing, but they give no interpretation about why the masses of three 
neutrinos are so tiny. A simple way to improve our phenomenological
ans$\rm\ddot{a}$tze is to incorporate them with the elegant idea 
of seesaw \cite{SS}. In the seesaw mechanism, 
the smallness of left-handed Majorana neutrinos is attributed to
the existence of heavy right-handed Majorana neutrinos,
\begin{equation}
M_\nu \; \approx \; M_{\rm D} M^{-1}_{\rm R} M^T_{\rm D} \; ,
\end{equation}
where $M_{\rm D}$ and $M_{\rm R}$ denote the Dirac neutrino mass
matrix and the heavy Majorana neutrino mass matrix, respectively.
In some grand unified theories (such as the SO(10) model \cite{SO10}), 
one takes $[M_{\rm D}, M_{\rm u}] = 0$, where $M_{\rm u}$ represents the 
up-type quark mass matrix. The mass matrix $M_{\rm R}$ is 
practically unknown in almost all reasonable extensions of the
standard model. Hence specific textures of $M_{\rm R}$ and 
$M_{\rm D}$ have to be assumed, in order to determine the texture
of $M_\nu$. Given $M_\nu$ and $M_{\rm D}$, on the other hand, one 
can calculate $M_{\rm R}$ by use of Eq. (25):
\begin{equation}
M_{\rm R} \; \approx \; M^T_{\rm D} M^{-1}_\nu M_{\rm D} \; .
\end{equation}
The scale of $M_{\rm R}$ stands for the scale of new physics in this
simple seesaw picture. 

To be specific, we assume that $M_{\rm D} = M_{\rm u}$ holds and
it has the same texture zeros as $M_l$ and $M_\nu$ have:
\begin{equation}
M_{\rm D} \; = \; \left ( \matrix{
{\bf 0} & C_{\rm u} & {\bf 0} \cr
C_{\rm u} & \tilde{B}_{\rm u} & B_{\rm u} \cr
{\bf 0} & B_{\rm u} & A_{\rm u} \cr} \right ) \; .
\end{equation}
Only if the condition ${\rm Det}M_\nu \neq 0$ is guaranteed
for $M_\nu$ in Eq. (2), one may obtain the inverse matrix of $M_\nu$
as follows:
\begin{equation}
M^{-1}_\nu \; =\; \frac{1}{A_\nu C^2_\nu} \left ( \matrix{
B^2_\nu - A_\nu \tilde{B}_\nu & A_\nu C_\nu & -B_\nu C_\nu \cr
A_\nu C_\nu & {\bf 0} & {\bf 0} \cr
-B_\nu C_\nu & {\bf 0} & C^2_\nu \cr} \right ) \; .
\end{equation}
Then the texture of $M_{\rm R}$ can be determined from Eq. (26) with
the help of Eqs. (27) and (28):
\begin{equation}
M_{\rm R} \; = \; \left ( \matrix{
{\bf 0} & C_{\rm R} & {\bf 0} \cr
C_{\rm R} & \tilde{B}_{\rm R} & B_{\rm R} \cr
{\bf 0} & B_{\rm R} & A_{\rm R} \cr} \right ) \; ,
\end{equation}
where
\begin{eqnarray}
A_{\rm R} & = & \frac{A_{\rm u}^2}{A_\nu} \; ,
\nonumber \\
B_{\rm R} & = & \frac{A_{\rm u} B_{\rm u}}{A_\nu}
~ + ~ \frac{B_{\rm u} C_{\rm u}}{C_\nu}
~ - ~ \frac{A_{\rm u} C_{\rm u} B_\nu}{A_\nu C_\nu} \; ,
\nonumber \\
\tilde{B}_{\rm R} & = & \frac{B_{\rm u}^2}{A_\nu}
~ + ~ \frac{2 \tilde{B}_{\rm u} C_{\rm u}}{C_\nu}
~ - ~ \frac{C_{\rm u}^2 \tilde{B}_\nu}{C^2_\nu}
~ - ~ \frac{2 B_{\rm u} C_{\rm u} B_\nu}{A_\nu C_\nu}
~ + ~ \frac{C_{\rm u}^2 B^2_\nu}{A_\nu C^2_\nu} \; ,
\nonumber \\
C_{\rm R} & = & \frac{C_{\rm u}^2}{C_\nu} \; .
\end{eqnarray}
We see that the texture zeros of $M_{\rm D}$ and $M_\nu$ manifest
themselves again in $M_{\rm R}$, as a consequence of the 
inverted seesaw relation given in Eq. (26). Therefore, all four 
lepton mass matrices ($M_l$, $M_{\rm D}$, $M_{\rm R}$ and $M_\nu$) 
are structurally parallel to one another. Such a structural 
similarity of lepton mass matrices, which is seesaw-invariant, 
might follow from a universal flavor symmetry hidden in a more 
fundamental theory of fermion mass generation. In particular,
the underlying flavor symmetry must be related to the texture
zeros of lepton mass matrices. It is worth mentioning that the
Fritzsch texture of lepton mass matrices in Eq. (1) does not have 
the interesting property of seesaw invariance. Thus we argue that
the four-zero texture of lepton mass matrix might be more attractive
for model building at the energy scale where the seesaw mechanism 
works. 

Now let us give an order-of-magnitude estimate of the matrix
elements of $M_{\rm R}$ by taking the following 
phenomenologically-favored pattern of $M_{\rm u}$ \cite{FX99}:
\begin{equation}
M_{\rm u} \; \sim \; \left ( \matrix{
{\bf 0}	& \sqrt{m_u m_c}	& {\bf 0} \cr
\sqrt{m_u m_c}	& m_c 	& \sqrt{m_u m_t} \cr
{\bf 0}	& \sqrt{m_u m_t}	& m_t \cr} \right ) \; ,
\end{equation}
in which the relevant complex phases have been neglected for 
simplicity. Typically taking $m_u/m_c \sim m_c/m_t \sim 0.0031$ and 
$m_t \approx 175$ GeV at the electroweak scale \cite{PDG}, we obtain  
\begin{equation}
M_{\rm R} \; \sim \; 8.8 \times 10^{14} ~ {\rm GeV} \times
\left ( \matrix{
{\bf 0} & 5.4 \times 10^{-8} & {\bf 0} \cr
5.4 \times 10^{-8} & 9.6 \times 10^{-6} & 3.1 \times 10^{-3} \cr
{\bf 0} & 3.1 \times 10^{-3} & {\bf 1} \cr} \right ) \; 
\end{equation}
from Eq. (19) for Ansatz (A); and
\begin{equation}
M_{\rm R} \; \sim \; 1.1 \times 10^{15} ~ {\rm GeV} \times
\left ( \matrix{
{\bf 0} & 7.3 \times 10^{-8} & {\bf 0} \cr
7.3 \times 10^{-8} & 9.6 \times 10^{-6} & 3.1 \times 10^{-3} \cr
{\bf 0} & 3.1 \times 10^{-3} & {\bf 1} \cr} \right ) \; 
\end{equation}
from Eq. (24) for Ansatz (B). We see that the structure of 
$M_{\rm R}$ is strongly hierarchical in either case. The scale of
$M_{\rm R}$ is about $10^{15}$ GeV, close to the typical scale of 
grand unified theories $\Lambda_{\rm GUT} \sim 10^{16}$ GeV.

It is worth remarking that the phase parameters of $M_{\rm D}$
and $M_\nu$ have been ignored in estimating the matrix elements
of $M_{\rm R}$. If those complex phases are included,
it is possible to get CP violation in the lepton-number-violating
decays of heavy Majorana neutrinos \cite{FY}. However, a
successful interpretation of the observed matter-antimatter 
asymmetry of the universe via the leptogenesis mechanism \cite{FY}
is rather nontrivial, because the details of $M_{\rm D}$ and 
$M_{\rm R}$ have to be taken into account. Further discussions on
this topic are interesting but beyond the scope of this paper.

\section{Summary}

We have proposed and discussed two phenomenological ans$\rm\ddot{a}$tze
of lepton mass matrices with four texture zeros. The parameter space
of each ansatz has been carefully analyzed by use of current experimental 
data on neutrino oscillations. We demonstrate that the normal hierarchy
of neutrino masses and the bi-large pattern of lepton flavor mixing 
can be accommodated in both ans$\rm\ddot{a}$tze. Their predictions for
the effective mass of the tritium beta decay and that of the 
neutrinoless double beta decay are too small to be detected in
practice. However, we find that leptonic CP violation at the percent
level is possible for either ansatz. 
The correlation of relevant observable quantities in each ansatz
allows us to test its validity, once more accurate experimental data
become available. This property may also allow us to distinguish between
these two different ans$\rm\ddot{a}$tze. 

For the purpose of illustration, we have presented some brief 
discussions about the seesaw realization of our phenomenological
scenarios. It is clear that the existence of heavy right-handed 
Majorana neutrinos at the scale of $10^{15}$ GeV or so 
may naturally interpret
the smallness of left-handed Majorana neutrino masses. This observation
would be useful for model building, from which some deeper understanding
of the neutrino mass generation and lepton flavor mixing could be gained.

One of us (H.Z.) is grateful to the theory division of IHEP for
financial support and hospitality in Beijing.
This work was supported in part by the National Natural Science 
Foundation of China.



\newpage

\begin{figure}[t]
\vspace{-2cm}
\epsfig{file=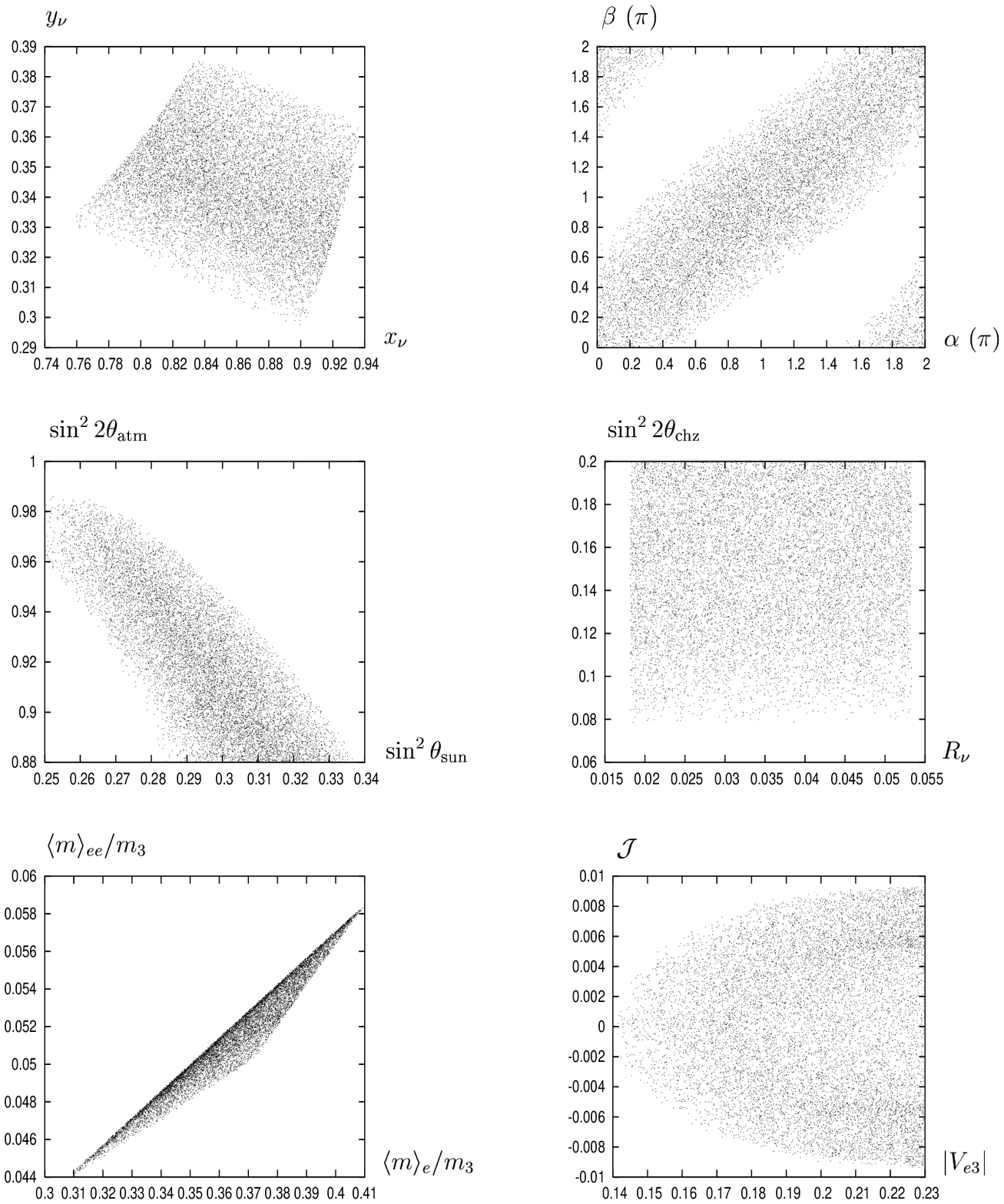,bbllx=1.7cm,bblly=5cm,bburx=19cm,bbury=29cm,%
width=16cm,height=24cm,angle=0,clip=0}
\vspace{-2cm}
\caption{The parameter space and phenomenological predictions of 
Ansatz (A).}
\end{figure}

\begin{figure}[t]
\vspace{-2cm}
\epsfig{file=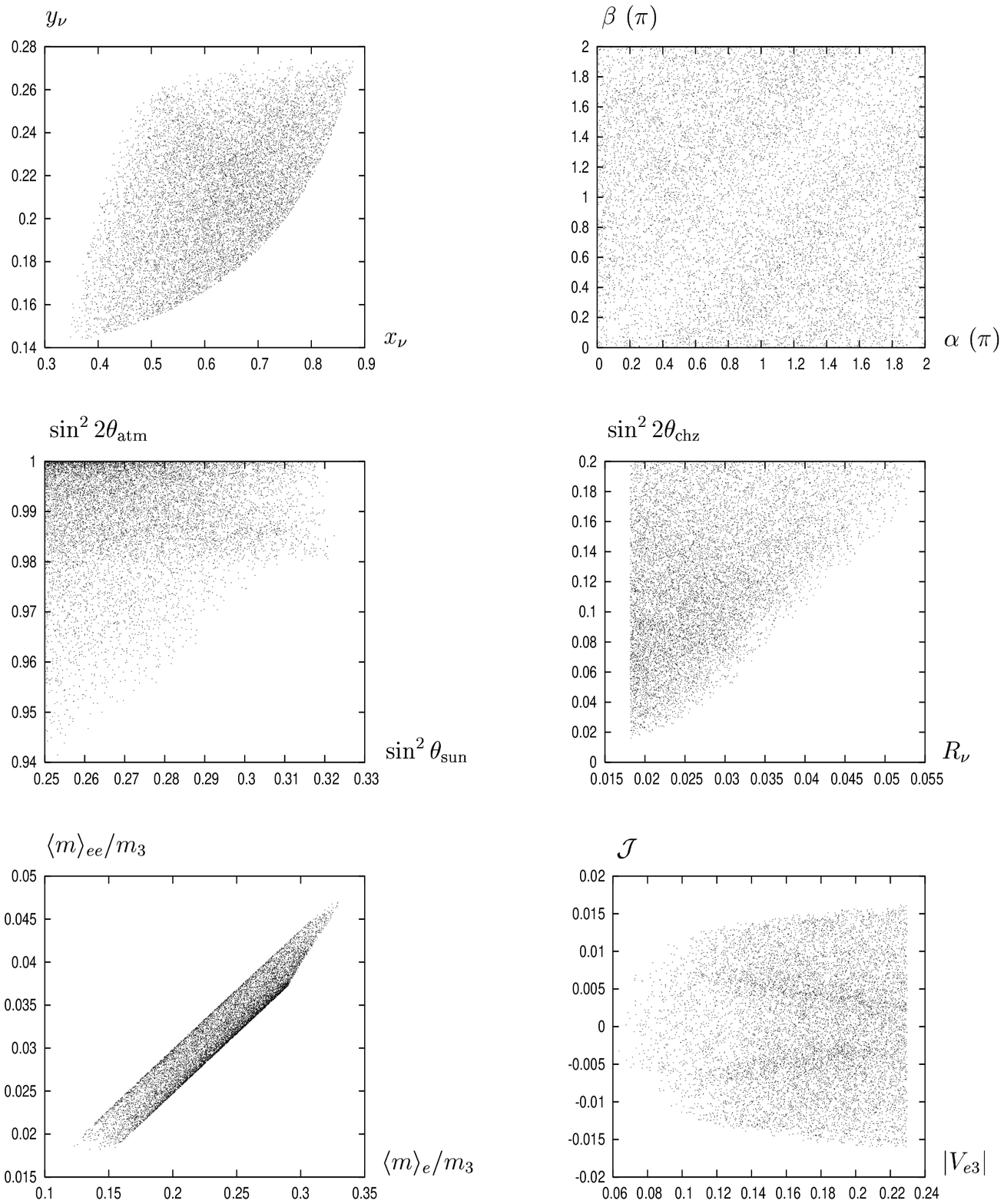,bbllx=1.7cm,bblly=5cm,bburx=19cm,bbury=29cm,%
width=16cm,height=24cm,angle=0,clip=0}
\vspace{-2cm}
\caption{The parameter space and phenomenological predictions of 
Ansatz (B).}
\end{figure}

\end{document}